%% file: hadron2011.tex
\documentclass[a4paper,11pt]{article}

\usepackage{contribution}

\input{econfmacros}
\input{contributionmacros}

\begin{document}

\input{contribution}

\end{document}

%% file: econfmacros.tex
\newcommand{\weblink}[2][]{%
    \ifthenelse{\equal{#1}{}}%
    {\textnormal{\url{#2}}}%
    {\textnormal{\href{#2}{#1}}}%
}

\def\beq{\begin{equation}}
\def\eeq#1{\label{#1}\end{equation}}
\def\eeqn{\end{equation}}

\def\beqa{\begin{eqnarray}}
\def\eeqa#1{\label{#1}\end{eqnarray}}
\def\eeqan{\end{eqnarray}}

\let\bar=\overbar

\def\Dslash{\not{\hbox{\kern-4pt $D$}}}
\def\dslash{\not{\hbox{\kern-2pt $\del$}}}

\def\msb{{\bar{\ssstyle M \kern -1pt S}}}

%% file: contributionmacros.tex
\newcommand{\contribution}[7][]{%
  \clearpage
  \thispagestyle{plain}
  \ifthenelse{\equal{#1}{}}
  {\hypersetup{pdftitle={#2}}}
  {\hypersetup{pdftitle={#1}}}
  \hypersetup{pdfauthor={{#3} {#4}}}
  {\centering\normalfont\LARGE\bfseries\sffamily #2 \par\nobreak}
  \lhead{}
  \chead{%
    \textit{\footnotesize XIV International Conference on Hadron Spectroscopy
      (\weblink[\textit{hadron2011}]{http://www.hadron2011.de}), 13-17 June 2011, Munich, Germany}%
  }
  \rhead{}
  \bigskip
  \begin{center}
    {#3} {#4}\ifthenelse{\equal{#6}{}}{}{\footnote{\weblink[#6]{mailto:#6}}}
    \ifthenelse{\equal{#7}{}}{}{#7} \\
    \textit{#5}
  \end{center}
  \bigskip
}

\renewcommand{\abstract}[1]{%
  \begin{center}
    \begin{minipage}{0.85\textwidth}
      \begin{footnotesize}
        #1
      \end{footnotesize}
    \end{minipage}
  \end{center}
  \bigskip
}

%% file: contribution.tex
%
%
%
%
%
{  

\def\psai{\psi(2S)}
\def\chic{\chi_c}
\def\mev{{\rm MeV}/c^2}
%

\contribution[Exotic spectroscopy and quarkonia at LHCb]  
{Exotic spectroscopy and quarkonia at LHCb}  
{Bo}{Liu}  
{Department of Engineering Physics,
  Tsinghua Unversity, \\
  100084 Beijing, CHINA \\
  and \\
  Laboratoire de l'Accelerateur Lineaire, Universite Paris-Sud 11, CNRS/IN2P3 (UMR 8607),\\
  91400 Orsay, FRANCE }  
{b-liu03@mails.thu.edu.cn}  
{on behalf of the LHCb Collaboration}  
%

\abstract{%
  Measurements of the X(3872) mass, $\psi(2S)$ production cross-section and
  $\chi_{c2}/\chi_{c1}$ production ratio
  using around 35$\rm pb^{-1}$ of pp collision data
  at $\sqrt{s}$ = 7 TeV during the 2010 LHC running period with the LHCb detector are presented.\\

}
%

\section{Introduction}
The LHCb detector is a single-arm forward spectrometer designed
to make precision measurements of CP violation and rare decays
in the heavy flavour at the LHC\cite{LHCb}. It has a unique geometrical
acceptance covering the pseudo-rapidity 1.9-4.9. This combined with excellent trigger, tracking and particle identification capabilities also makes it an ideal detector to study quarkonia.\\
This paper presents the measurements of the X(3872) mass, $\psi(2S)$ production cross-section and
  $\chi_c$ production ratio in pp collisions at $\sqrt{s}$ = 7 TeV with the data taken by the LHCb
detector in 2010.

\section{Measurement of X(3872) mass}

The X(3872) meson is an exotic meson discovered in 2003 by the Belle collaboration\cite{belle}.
Several properties of the X(3872) have been measured,
but its nature is still uncertain
and several models have been proposed. First, it is not excluded that the X(3872) is a
conventional charmonium state with one candidate being the $\eta_{c2}(1D)$ meson\cite{xconf5}. However,
the predicted mass of this state is far below the observed X(3872) mass. Given the
proximity of the X(3872) mass to the $D^{*0}\bar{D}^0$ threshold,
one possibility is that the X(3872)
is a loosely bound deuteron-like $D^{*0}\bar{D}^0$ molecule,
i.e. a $((u\bar{c}) - (c\bar{u}))$ system\cite{xconf5}. Another
more exotic possibility is that the X(3872) is a tetraquark state\cite{xconf8}.
A precise measurement of the mass is an important ingredient towards elucidating the nature of this state.


For accurate measurement of particle mass systematic uncertainties
due to the momentum scale must be under control.
The momentum scale\cite{xconf}\cite{massconf} is calibrated using a large sample of $J/\psi\rightarrow\mu^+\mu^-$ decays. This
calibration accounts for a mixture of effects related to imperfect
knowledge of the magnetic field map and of the alignment of the tracking system.
The calibration is cross-checked using two-body decay of the $\Upsilon(1S)$, $K^0_S$ and $D^0$.
Based on comparisions of the values obtained for these masses and those quoted
by the PDG \cite{PDG} a systematic uncertainty of 0.1 per mille is assigned.


%
The X(3872) mass measurement is affected by the systematic uncertainties from the mass fitting
(signal model (0.02 $\mev$), background model (0.02 $\mev$)), average momentum scale (0.05 $\mev$)
, pseudo-rapidity dependence of momentum scale (0.03 $\mev$),
energy loss correction of the detector description (0.05 $\mev$),
alignment of tracking stations (0.05 $\mev$) and
alignment of vertex detector (0.01 $\mev$). The total
systematic uncertainty is 0.10 $\mev$.


Fig. ~\ref{fig:xmass} shows the the invariant $J/\psi\pi^+\pi^-$ mass distributions for opposite-sign (black points)
 and same-sign (blue filled histogram) candidates. Clear signals for both the $\psai$ and the
X(3872) can be seen. The preliminary result is $M_{X(3872)}=3871.96\pm0.46(\rm stat)\pm0.10(\rm syst) {\rm MeV}/c^2$.
\begin{figure}[htb]
  \begin{center}
    \includegraphics[width=0.5\textwidth]{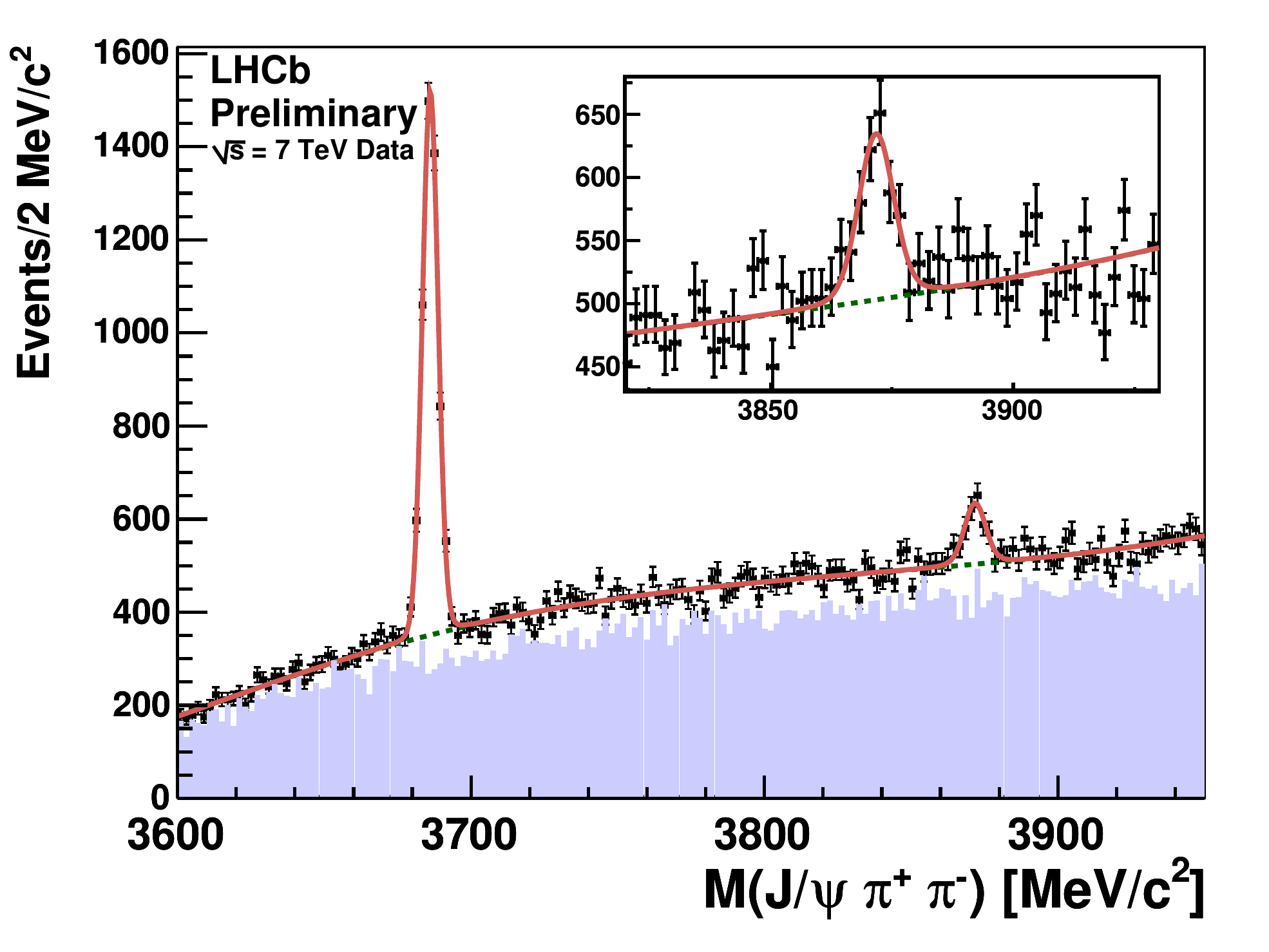}
    \caption{Invariant mass distribution of $J/\psi\pi^+\pi^-$ (black points with statistical error
bars) and same-sign $J/\psi\pi^\pm\pi^\pm$ (blue filled histogram) candidates. The red curve is the
result of the fit to the mass distribution. The insert shows a zoom of the region around the
X(3872) mass.}
    \label{fig:xmass}
  \end{center}
\end{figure}
%

%

\section{Measurement of $\psi(2S)$ cross-section}

Prompt heavy quarkonium production in hadron collisions is a subject of large interest\cite{psithe}.
In order to ensure an unambiguous comparison with theory, promptly produced
quarkonia should be separated from those coming from b-hadron decays and from those cascading from higher mass states (feed-down): the $\psi(2S)$ meson has no appreciable feed-down
from higher mass states and therefore the results can be directly compared with the theory,
making it an ideal laboratory for QCD studies.

The differential cross-section\cite{psiconf}
 for the inclusive $\psai$ production is
 computed as
$$\frac{d^2\sigma}{dp_{\rm T}}(p_{\rm T})=\frac{N_{\psai}(p_{\rm T})}
{L_{\rm int}\epsilon(p_{\rm T})B(\psai\rightarrow e^+e^-)\Delta p_{\rm T}}$$
where $N_\psai(p_{\rm T})$ is the number of observed $\psai\rightarrow J/\psi\pi^+\pi^-$ decays,
$L_{\rm int}$ is the integrated
luminosity,
$\epsilon(p_{\rm T})$ is the total detection efficiency,
$B(\psai\rightarrow J/\psi\pi^+\pi^-)$ is the $\psai\rightarrow J/\psi\pi^+\pi^-$
branching ratio\cite{PDG}
and $\Delta p_{\rm T}$ is the bin size.
 In order to estimate
the number of $\psai$ signal events, a fit to the $\psai$ mass distribution is performed independently
in each $p_{\rm T}$ bin.

The total efficiency $\epsilon$ is the product of the
geometric acceptance (A), the reconstruction efficiency ($\epsilon_{\rm rec}$) and the trigger efficiency ($\epsilon_{\rm tri}$).
The factor $\epsilon_{\rm rec}$ includes the detection, reconstruction and selection efficiency. Monte Carlo
simulations are used to estimate A, $\epsilon_{\rm rec}$ and $\epsilon_{\rm tri}$ for each bin.
The dominant systematic uncertainties are from the luminosity measurement 10\%,
the unknown polarization (bin-dependent), the trigger efficiency (bin-dependent) and
the tracking efficiency (4\% per track).


  The integrated cross-section in the full range of $p_{\rm T}$ and rapidity ($y$) is found to be
 $$\sigma(3<p_{\rm T}\leq16{\rm GeV}/c, 2<y\leq4.5) = 0.62\pm0.04\pm0.12^{+0.07}_{-0.14}{\rm \mu b}$$
  where the first error is statistical,
   the second error is systematic and the third asymmetric error
 is related to the unknown polarization.

    The double differential cross-sections for the $\psai\rightarrow J/\psi\pi^+\pi^-$ mode,
     another measurement
    of $\psai\rightarrow\mu^+\mu^-$ mode\cite{psiconf}
     and comparison
    with the prediction of prompt production by a NLO NRQCD model\cite{psicomp}
     are shown in Fig. ~\ref{fig:psixs}.
     The results for the two decay modes and the theory are in good agreement.
\begin{figure}[htb]
  \begin{center}
    \includegraphics[width=0.5\textwidth]{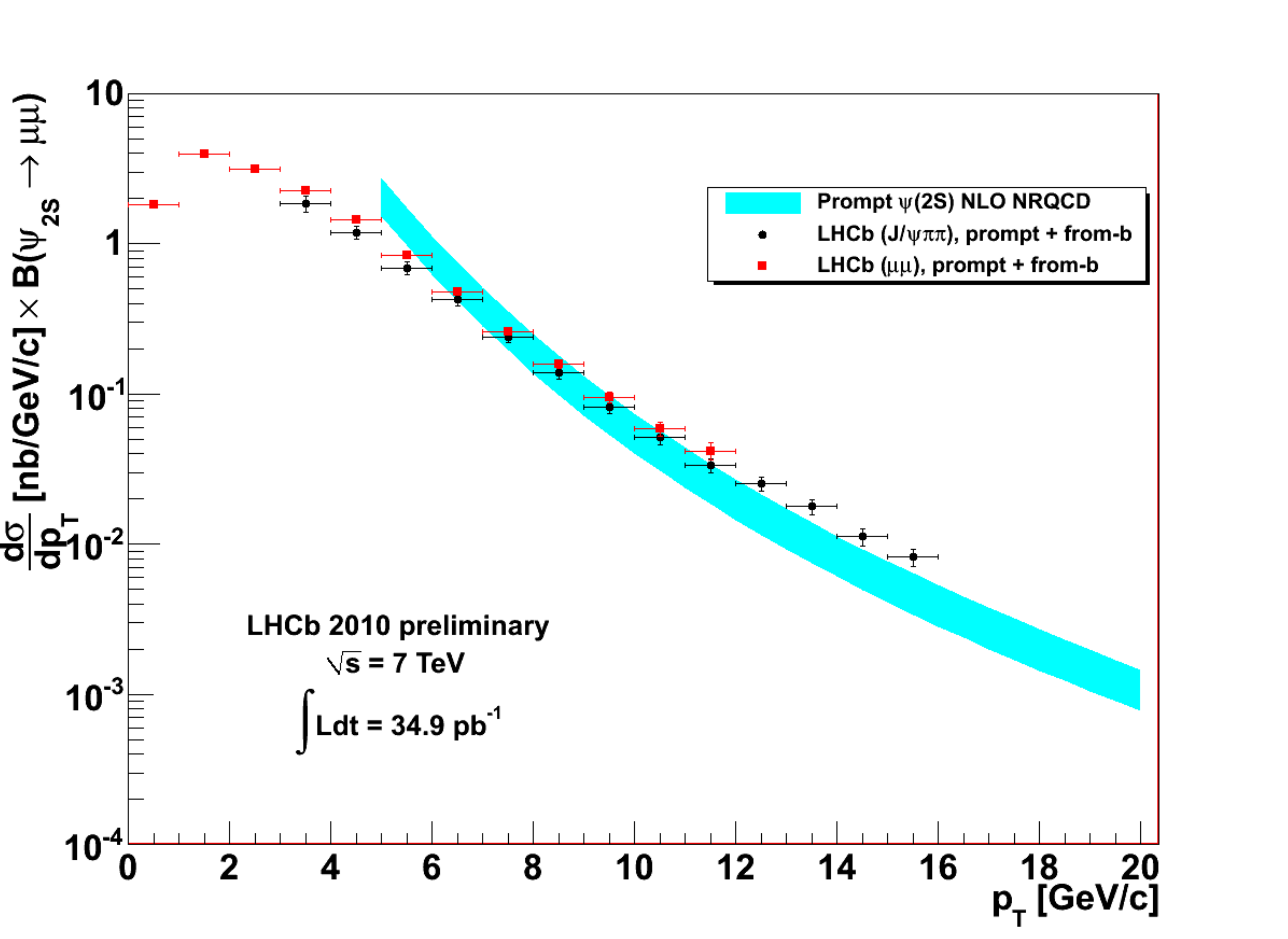}
    \caption{Comparison of the LHCb results for the differential production cross-section of $\psai$
with the predictions for prompt production by a NLO NRQCD model\cite{chiccomp}. LHCb data include
also $\psai$ from b: the fraction of $\psai$ from b is expected to be of the order of 10$\%$ at low $p_{\rm T}$
and to increase as function of $p_{\rm T}$ to about 40$\%$\cite{psinum}.}
    \label{fig:psixs}
  \end{center}
\end{figure}
%

\section{Measurement of $\chi_c$ cross-section ratio}
The study of P-wave charmonia $\chic$(J=0,1,2)
production is important for two reasons.
First, the measurement of the production rates of $\chi_{c2}$ and $\chi_{c1}$ is sensitive
to the Color-Singlet and Color-Octet production mechanisms. In addition, the feeddown contribution to prompt $J/\psi$ production through radiative decays has important consequences for
the $J/\psi$ polarization measurement\cite{chicconf}.

The production cross-section ratio of the $\chi_{c2}$ and $\chi_{c1}$ states is measured using
$$\frac{\sigma(\chi_{c2})}{\sigma(\chi_{c1})}=
\frac{N_{\chi_{c2}}}{N_{\chi_{c1}}}
\times\frac{\epsilon^{\chi_{c1}}_{J/\psi}\epsilon^{\chi_{c1}}_{\gamma}\epsilon^{\chi_{c1}}_{\rm sel}}
{\epsilon^{\chi_{c2}}_{J/\psi}\epsilon^{\chi_{c2}}_{\gamma}\epsilon^{\chi_{c2}}_{\rm sel}}
\times\frac{B(\chi_{c1}\rightarrow J/\psi\gamma)}{B(\chi_{c2}\rightarrow J/\psi\gamma)}$$
where $B(\chi_{c1}\rightarrow J/\psi\gamma)$($B(\chi_{c2}\rightarrow J/\psi\gamma)$)
are the $\chi_{c1}$($\chi_{c2}$) branching ratios to the final state $J/\psi\gamma$.
The event yield $N_{\chi_c}$ is from an unbinned maximum likelihood fit to the mass distribution.
 The efficiencies $\epsilon$
 are extracted
from the Monte Carlo simulation.

%

The main systematic uncertainties are from the fit (the fit model, fit range and fit method),
the MC statistics, the unknown polarization and the $\chic\rightarrow J/\psi\gamma$ branching ratios.

The preliminary result for the ratio of the prompt $\chi_{c2}$ to $\chi_{c1}$ production cross-sections,$\frac{\sigma{\chi_{c2}}}{\sigma{\chi_{c1}}}$, as a function of $p^{J/\psi}_{\rm T}$ is shown in Figure ~\ref{fig:chicxs}.
 The preliminary result is broadly in agreement with the theorectical
 predictions at high $p^{J/\psi}_{\rm T}$. At low $p^{J/\psi}_{\rm T}$ there are indications of a discrepancy.

\begin{figure}[htb]
  \begin{center}
    \includegraphics[width=0.5\textwidth]{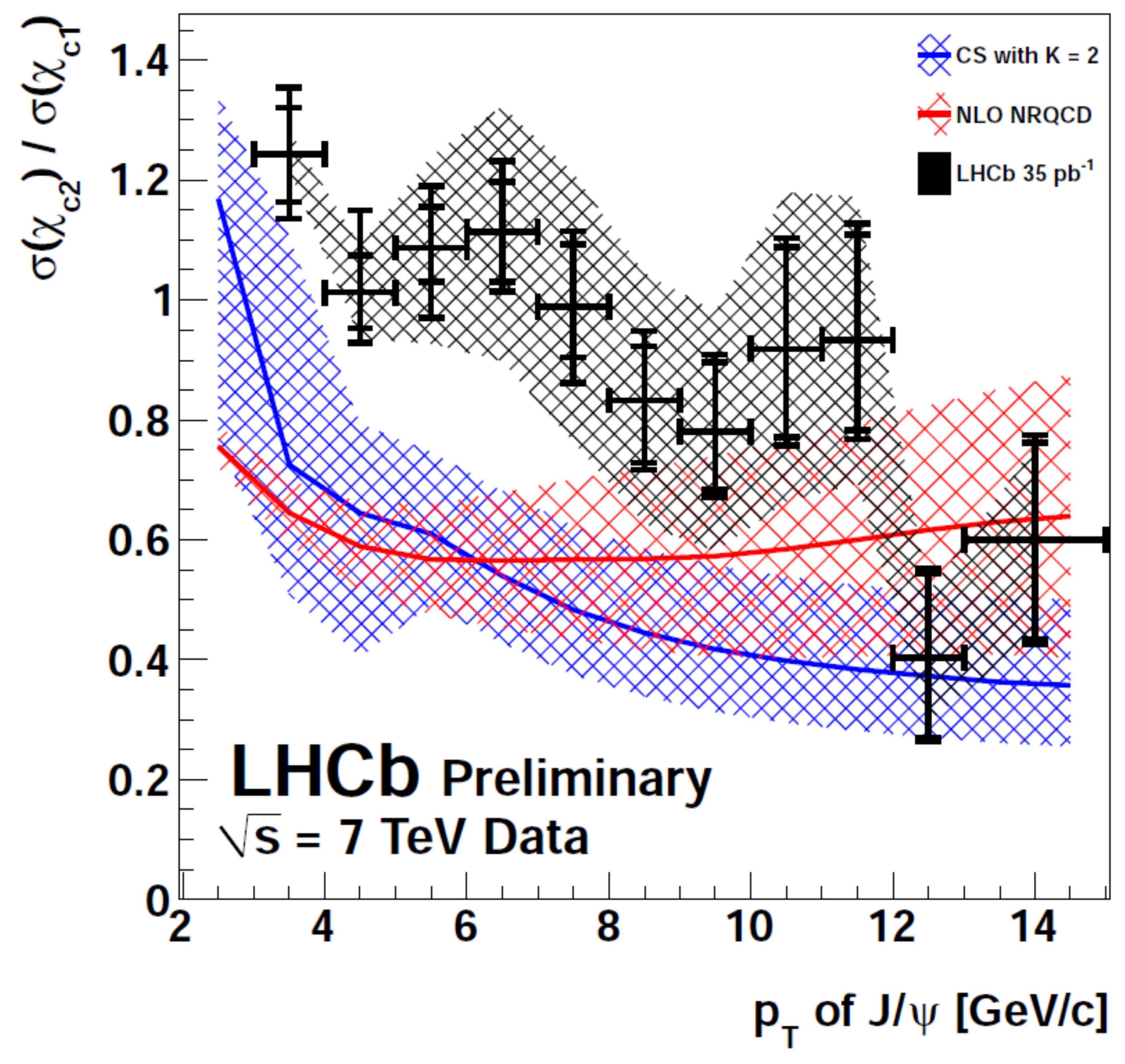}
    \caption{The ratio $\frac{\sigma{\chi_{c2}}}{\sigma{\chi_{c1}}}$ in bins of $p^{J/\psi}_{\rm T}$ $\in$ [3; 15]GeV/c. The internal error bars correspond
to the statistical error on the $\chi_{c1}$ and $\chi_{c2}$ yields; the external error bars include the contribution
from all the systematic uncertainties. The shaded area around the data points (black) shows
the maximum effect of the unknown $\chic$ polarizations on the result. The upper limit corresponds
to the spin state ($\chi_{c1}$: $m_J$ = 1; $\chi_{c2}$: $m_J$ = 2) and the lower limit corresponds to the spin state
($\chi_{c1}$: $m_J$ = 0; $\chi_{c2}$: $m_J$ = 0). The two other bands correspond to the ChiGen MC generator
theoretical prediction \cite{chicmc}(in blue) and NLO NRQCD (in red)\cite{chiccomp}.}
    \label{fig:chicxs}
  \end{center}
\end{figure}
%

\section{Summary}


First results from studies of X(3872), $\chi_c$ and $\psi(2S)$ mesons using data collected
 by the LHCb detector have been presented. The results for the cross-sections
 are in reasonable agreement with theory.
 The larger data-set collected during the 2011 running period will allow the statistical
uncertainties to be reduced and further studies (for example of the $\psi(2S)$ polarization) will be performed.




}  
